\documentclass[letter]{ieice}
\usepackage{graphicx}

\usepackage{threeparttable,bm,amsmath,graphicx,amssymb,comment,array,multirow,threeparttable,cite,calligra,color}
\field{}
\title{Security Evaluation for Block Scrambling-Based Image Encryption Including JPEG Distortion against Jigsaw Puzzle Solver Attacks}
\authorlist{
\authorentry[chuman-tatsuya@ed.tmu.ac.jp]{Tatsuya CHUMAN}{n}{tmu}\MembershipNumber{}
\authorentry[kiya@tmu.ac.jp]{Hitoshi KIYA}{f}{tmu}\MembershipNumber{}}
\affiliate[tmu]{The authors are with Tokyo Metropolitan University, Hino-shi, 191-0065 Japan}
\received{2015}{1}{1}
\revised{2015}{1}{1}

\begin{document}
\maketitle
\setlength{\abovedisplayskip}{4pt}
\begin{summary}
Encryption-then-Compression (EtC) systems have been considered for the user-controllable privacy protection of social media like Twitter.
The aim of this paper is to evaluate the security of block scrambling-based encryption schemes, which have been proposed to construct EtC systems.
Even though this scheme has enough key spaces against brute-force attacks, each block in encrypted images has almost the same correlation as that of original images.
Therefore, it is required to consider the security from different viewpoints from number theory-based encryption methods with provable security such as RSA and AES.
In this paper, we evaluate the security of encrypted images including JPEG distortion by using automatic jigsaw puzzle solvers.
\end{summary}


\section{Introduction}
The use of images and video sequences has greatly increased because of the rapid growth of the Internet and widespread use of multimedia systems.
While many studies on secure, efficient, and flexible communications have been reported\cite{huang2014survey,lagendijk2013encrypted}, full encryption with provable security (like RSA and AES) is the most secure option for securing multimedia data.
However, there is a trade-off between security and other requirements such as low processing demand, bitstream compliance, and signal processing in the encrypted domain. Several perceptual encryption schemes have been developed to achieve this trade-off\cite{Tang_2014,Chengqing}.
\par
In this paper, we focus on block scrambling-based image encryption schemes, which have been proposed for Encryption-then-Compression (EtC) systems with the assumption of international compression standards to consider the safety\cite{watanabe2015encryption,KURIHARA2015,Kuri_2017}.
So far, the safety has been evaluated based on its key space assuming the brute-force attacks, so that the schemes have enough key spaces for protecting the attacks.
However, each block in encrypted images has almost the same correlation as that of original images. Several efficient attacks on the permutation-only encryption have been studied\cite{jolfaei2016security}, but they are not available for the block scrambling-based encryption.
\par
On the other hand, recently, jigsaw puzzle solvers, that utilize the correlation between pieces, have succeeded in solving puzzles with a large number of pieces\cite{Gallagher_2012_CVPR, Cho_2010_CVPR, Sholomon_2016_GPEM, Rui_2015_ArXiv}.
Furthermore, regarding the blocks of an encrypted image as pieces of a jigsaw puzzle, the new types of jigsaw puzzle solvers for the attacks have been proposed\cite{CHUMAN2017ICASSP,CHUMAN2017ICME,CHUMAN2017IEICE}.
However, these methods do not consider encrypted images including JPEG distortion made through Social Networking Services (SNS) providers.
In this paper, we utilize these puzzle solvers to evaluate the safety of the encrypted images including JPEG distortion.
\par
Finally, we evaluate the safety of the encryption by applying the jigsaw puzzle solvers to encrypted images with JPEG artifact on the assumption that JPEG standard was used for EtC systems.
It is shown that some solvers can assemble encrypted images partly even when the key space is large enough.
On the other hand, it is also confirmed that JPEG distortion makes the decryption of encrypted images more difficult than images with no distortion.

\section{Preparation}
\subsection{Block Scrambling-based Image Encryption}

\label{subsec:block}
\begin{figure}[t]
\centering
\includegraphics[width =7.2cm]{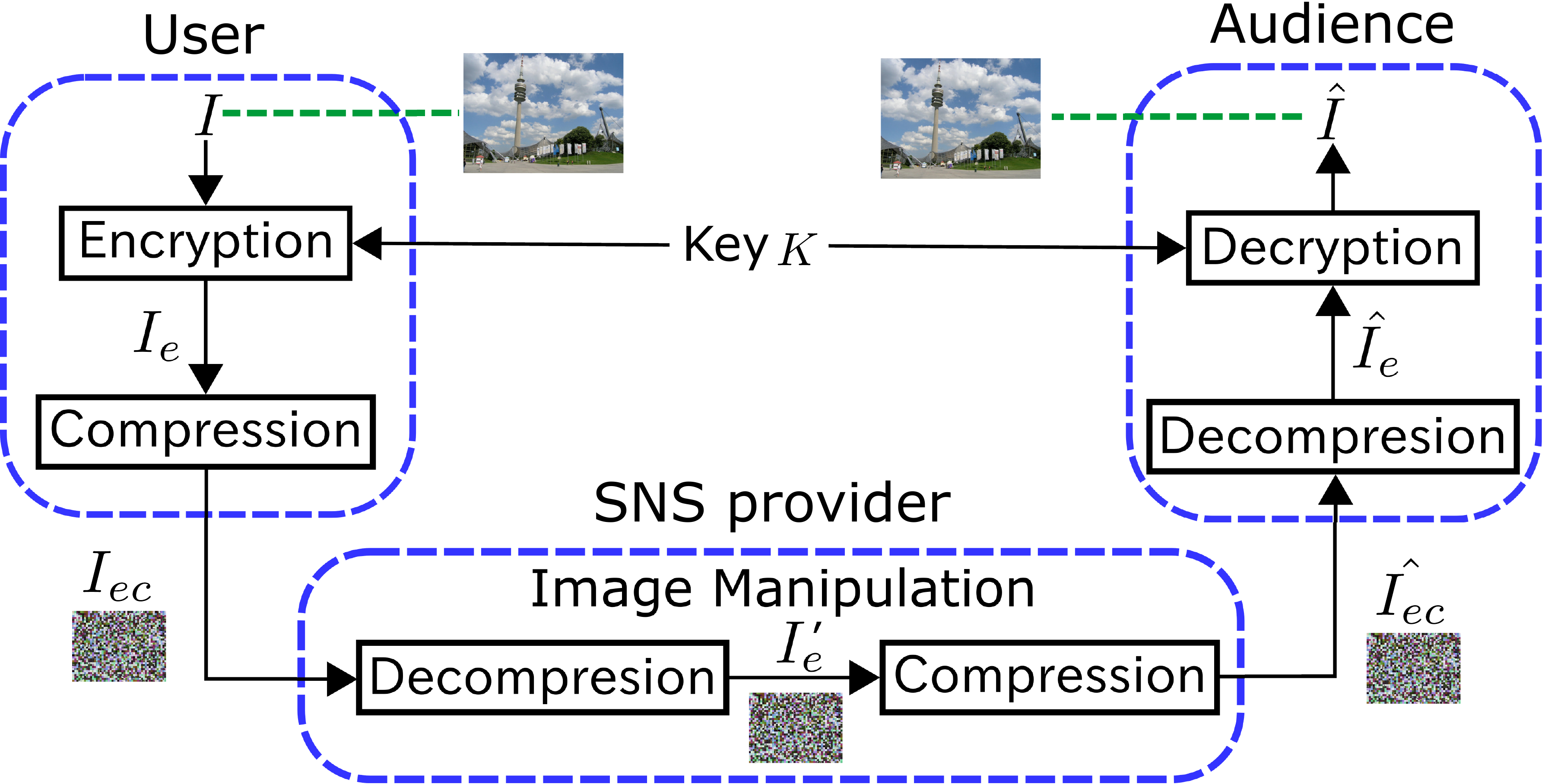}
\caption{Encryption-then-Compression system}
\label{fig:etc}
\vspace{-2mm}
\end{figure}

Block scrambling-based image encryption schemes have been proposed for EtC systems\cite{KURIHARA2015,Kuri_2017}, in which a user wants to securely transmit image $I$ to an audience, via a SNS provider, as illustrated in Fig.\,\ref{fig:etc}.
Since the user does not give the secret key $K$ to the SNS provider, the privacy of image to be shared is under control of the user even when the SNS provider recompresses image $I$.
Therefore, the user is able to protect the privacy by him/herself.
\par

\begin{figure}[t]
\centering
\includegraphics[width =8.4cm]{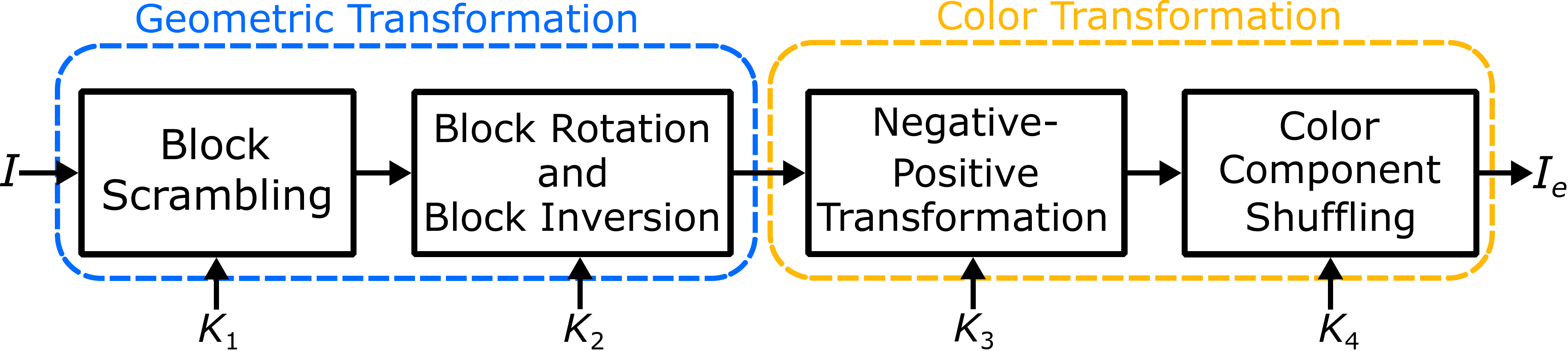}
\caption{Block scrambling-based image encryption}
\vspace{-1mm}
\label{fig:step}
\end{figure}

In the schemes\cite{watanabe2015encryption,KURIHARA2015,Kuri_2017}, an image with $X \times Y$ pixels is first divided into non-overlapped blocks with $B_x \times B_y$, then four block scrambling-based processing steps, as illustrated in Fig.\,\ref{fig:step}, is applied to the divided image.
The procedure of performing the image encryption to generate an encrypted image $I_e$ is given as follows:

\begin{itemize}
\setlength{\parskip}{0cm}
\setlength{\itemsep}{0cm}

\item[Step1:]
Divide an image with $X \times Y$ pixels into blocks with $B_x \times B_y$ pixels, and permute randomly the divided blocks using a random integer generated by a secret key $K_1$, where $K_1$ is commonly used for all color components.
Thus, the number of blocks $n$ is given by
\vspace{-1.5mm}
\begin{equation}
n = \lfloor \frac{X}{B_x} \rfloor \times \lfloor \frac{Y}{B_y} \rfloor
\end{equation}
where $\lfloor \cdot \rfloor$ is the function that rounds down to the nearest integer.

\item[Step2:]
Rotate and invert randomly each block using a random integer generated by a key $K_2$, where $K_2$ is commonly used for all color components as well.
\item[Step3:]
Apply the negative-positive transformation to each block using a random binary integer generated by a key $K_3$, where $K_3$ is commonly used for all color components. In this step, a transformed pixel value in $i$th block $B_i$, $p'$ is computed by
\vspace{-1mm}
\begin{equation}
p'=
\left\{
\begin{array}{ll}
p & (r(i)=0) \\
p \oplus (2^L-1) & (r(i)=1)
\end{array}
\right.
\vspace{-1mm}
\end{equation}
where $r(i)$ is a random binary integer generated by $K_3$ and $p \in B_i$ is the pixel value of an original image with $L$ bpp.
{In this paper, the value of occurrence probability $P(r(i))$=0.5 has been used to invert bits randomly.}

 \begin{table}[t]
 \caption{Permutation of color components for a random integer}
 \label{table:rgbcomponenttable}
 \begin{center}
     \begin{threeparttable}
        \scalebox{0.9}{
  \begin{tabular}{c|c|c|c|c}
   \hline
   Random Integer & R & G & B &Transform Function$f$ \\
   \hline
0 & R & G & B &$f_{RGB}$\\
1 & G & R & B &$f_{GRB}$\\
2 & R & B & G &$f_{RBG}$\\
3 & B & G & R &$f_{BGR}$\\
4 & B & R & G &$f_{BRG}$\\
5 & G & B & R &$f_{GBR}$\\
   \hline
  \end{tabular}
  }
\end{threeparttable}
\end{center}
\vspace{-5mm}
\end{table}

\item[Step4:]
Shuffle three color components in each block (the color component shuffling) using a random senary integer generated by a key $K_4$.
Table\,\ref{table:rgbcomponenttable} shows the permutation of color components corresponding to the random integer.
\end{itemize}

\subsection{Encrypted Images Including JPEG Distortion}
As shown in Fig.\,\ref{fig:etc}, a SNS provider receives an encrypted image $I_{ec}$ compressed by a user.
The decompressed image $I_{e}$ includes JPEG artifact if the user utilizes the JPEG standard as the method of compression.
Moreover, the SNS provider sends a recompressed image $I_{ec}$ to an audience.
As a result, the audience gets image $\hat{I}$ including JPEG distortion, generated by two JPEG operations.
{Note that the audience can decrypt ones with keys after decompressed images, even when the encrypted images are lossy compressed by the JPEG standard.}
\par
The key space of the block scrambling-based image encryption is generally large enough against the brute-force attacks\cite{KURIHARA2015}.
However, an encrypted image has almost the same correlation among pixels in each block as that of the original image, whose property enables to efficiently compress images.
Therefore, when a SNS provider leaks encrypted images such as $I_{e}$, an attacker can utilize the correlation to decrypt the image in some way.
The aim of this paper is to discuss the security of the encryption against jigsaw puzzle solver attacks that are based on the correlation under the condition that encrypted images include JPEG distortion.

\section{Extended Jigsaw Puzzles Solver}
Jigsaw puzzle solver is a method of assembling jigsaw puzzles. In the block scrambling-based encryption, if we regard the blocks as pieces of a jigsaw puzzle, decrypting encrypted images is similar to assembling the jigsaw puzzle. Therefore, jigsaw puzzle solvers are considered as one of the attack methods on the block scrambling-based encryption in this paper.

\subsection{Related Works}
Jigsaw puzzle solvers are broadly classified into three categories according to their assembly strategies, i.e., greedy methods, global methods and their hybrid methods\cite{Rui_2015_ArXiv}. The greedy methods start from initial pairwise matches and successfully build larger and larger components\cite{Cho_2010_CVPR,Gallagher_2012_CVPR}. On the other hand, the global methods directly search for a solution by maximizing a global compatibility function\cite{Sholomon_2016_GPEM}.
\par
The jigsaw puzzle solver\cite{Sholomon_2016_GPEM} completely succeeded in assembling large puzzles which consist of 30745 pieces with the size of $28\times28$, in 2016.
On the other hand, a solver for puzzles including rotated pieces (pieces with unknown orientation) was first proposed in 2012\cite{Gallagher_2012_CVPR}.
However, these puzzle solvers are available only for limited jigsaw puzzles which consist of pieces include just scrambled pieces and rotated ones.
\par
In this paper, jigsaw puzzle solvers are considered as one of attacks on the image encryption.
The existing jigsaw solvers do not support inverted, color component shuffled or negative-positive transformed pieces as mentioned above\cite{Gallagher_2012_CVPR, Cho_2010_CVPR, Sholomon_2016_GPEM, Rui_2015_ArXiv}.

\begingroup
\newcolumntype{A}{>{\centering\arraybackslash}p{1.38cm}}
\newcolumntype{B}{>{\centering\arraybackslash}p{1.65cm}}
\newcolumntype{C}{>{\centering\arraybackslash}p{1.3cm}}
\newcolumntype{D}{>{\centering\arraybackslash}p{1.25cm}}
\newcolumntype{E}{>{\centering\arraybackslash}p{2.2cm}}
\begin{table}[t]
\caption{Jigsaw puzzle types}
\begin{center}
\scalebox{0.71}{
\begin{threeparttable}
\label{tb:encryptionTable}
\begin{tabular}{A|C|D|D|E|B}
 \multirow{3}{*}{Type} & \multirow{3}{*}{\shortstack{Scramble}} & \multirow{3}{*}{\shortstack{Rotation}} &   \multirow{3}{*}{Inversion} &
 Negative-Positive Transformation & Color Component Shuffling\\
 \hline Type 1& \checkmark &  &  & &\\
 Type 2  & \checkmark & \checkmark &            &            &\\
 Type I  & \checkmark & \checkmark & \checkmark &            &\\
 Type N  & \checkmark & \checkmark &            & \checkmark &\\
 Type IN & \checkmark & \checkmark & \checkmark & \checkmark &\\
 Type INC& \checkmark & \checkmark & \checkmark & \checkmark & \checkmark
\end{tabular}
\end{threeparttable}}
\vspace{-5mm}
\end{center}
\end{table}
\endgroup

Therefore, we utilize the extended jigsaw puzzle solver\cite{CHUMAN2017ICASSP,CHUMAN2017ICME}, which enable to assemble Type I, N, IN and Type INC puzzles, where these types of jigsaw puzzles are indicated in Table\,\ref{tb:encryptionTable}.

\subsection{Extended Jigsaw Puzzle Solver}
\label{sec:new}
The extended jigsaw puzzle solver\cite{CHUMAN2017ICASSP} was proposed based on the greedy method\cite{Gallagher_2012_CVPR} to assemble jigsaw puzzles including inverted pieces, negative-positive transformed ones or component shuffled ones.
The following is the procedure.

\subsubsection{Pairwise Compatibility}
To calculate pairwise compatibility between pieces, we use Mahalanobis Gradient Compatibility (MGC) proposed by Gallagher\cite{Gallagher_2012_CVPR}.
Given the pieces $x_{i}$ and $x_{j}$, $i, j = 1, 2,\ldots,n$, the compatibility between the right side of $x_{i}$ and the left side of $x_{j}$ is expressed as $C_{LR}(x_{i}, x_{j} )$.

\subsubsection{Pairwise Comparison}
We represent transform function that rotates $x_{j}$ $0^\circ$, $90^\circ$, $180^\circ$ or $270^\circ$ as $f_{R}$, $R\in\{0,90,180,270\}$.
The function that inverts $x_{j}$ horizontally(H) or vertically(V) is defined as $f_{I}(x_{j})$, $I\in\{H,V,0\}$, where $f_{0}(x_{j})$ is the function that indicates non-inverted.
$f_{N}(x_{j})$, $N\in\{N,0\}$ is the function whether applies negative-positive transformation(N) to $x_{j}$.
In accordance with Table\,\ref{table:rgbcomponenttable}, the function that applies $x_{j}$ to color component shuffling is given as $f_{C}(x_{j})$, $C\in\{RGB$$,GRB$,$RBG$,$BGR$,$BRG$,$GBR$$\}$. 
In addition to four transform functions, i.e., $f_{R}(x_{j})$, $f_{I}(x_{j})$, $f_{N}(x_{j})$ and $f_{C}(x_{j})$, the combination of them gives other transformations.
Then, a rotated, inverted, negative-positive transformed and color component shuffled piece is represented as
 \begin{equation}
	\begin{array}{ll}
	f_{R,I,N,C}(x_{j}) =f_{R} \circ f_{I} \circ f_{N} \circ f_{C}(x_{j})
	\end{array}
\end{equation}
where $f_{R,I,N,C}(x_{j})$ is the composite function which consists of four transform functions.
\par
In the extended solver, the minimum compatibility between the right side of $x_{i}$ and the left side of $x_{j}$ is defined by
\begin{equation}
	\begin{array}{ll}
	\!\!\!\!\!\!\!\!\!\!\!\!\!\min{}C_{LR}(x_{i},x_{j}) \!=\!\!\!\! {\displaystyle \min_{f_{R,I,N,C}}\{C_{LR}(x_{i},f_{R,I,N,C}(x_{j}))}\}.
	\end{array}
\end{equation}
Finally, these minimum compatibility values are used to assemble jigsaw puzzle by using tree-based assembly method\cite{Gallagher_2012_CVPR}.

\section{Experiments and Results}
\subsection{Experimental Conditions}
\label{sec:gg}
Image $I_d$ assembled by jigsaw puzzle solvers from a Type I, N, IN or Type INC puzzle was compared with the original image $I$.
The following three measures\cite{Gallagher_2012_CVPR}\cite{Cho_2010_CVPR} were used to evaluate the results.
\\
{\bf Direct comparison ($Dc$):} represents the ratio of the number of pieces which are in the correct position.
\\
{\bf Neighbor comparison ($Nc$):} is the ratio of the number of correctly joined blocks.
\\
{\bf Largest component ($Lc$):} is the ratio of the number of the largest joined blocks which have correct adjacencies to the number of blocks in an image.
\par
In the measures, $Dc(I_d), Nc(I_d),Lc(I_d) \in [0,1]$, a larger value means a higher compatibility.
\par
We used 20 images from MIT dataset, provided by Cho\cite{Cho_2010_CVPR}.
Three different encrypted images were generated by random keys from one ordinary image for each Type puzzle{($B_{x}=B_{y}=32$)}.
Each encrypted image was compressed to add JPEG artifact by the encoder from the Independent JPEG Group (IJG) software\cite{JPEGLIB}.
Then, we assembled the encrypted images by using jigsaw puzzle solvers and chose the image which had the highest sum of $Dc(I_d)$, $Nc(I_d)$ and $Lc(I_d)$ in those of three images.
We performed these procedures for each type puzzle independently, and the average of 20 images was calculated for $Dc(I_d)$, $Nc(I_d)$ and $Lc(I_d)$.

\begin{figure*}[t]
   \begin{center}
      \begin{minipage}{0.33\hsize}
        \begin{center}
          \includegraphics[width=5.82cm]{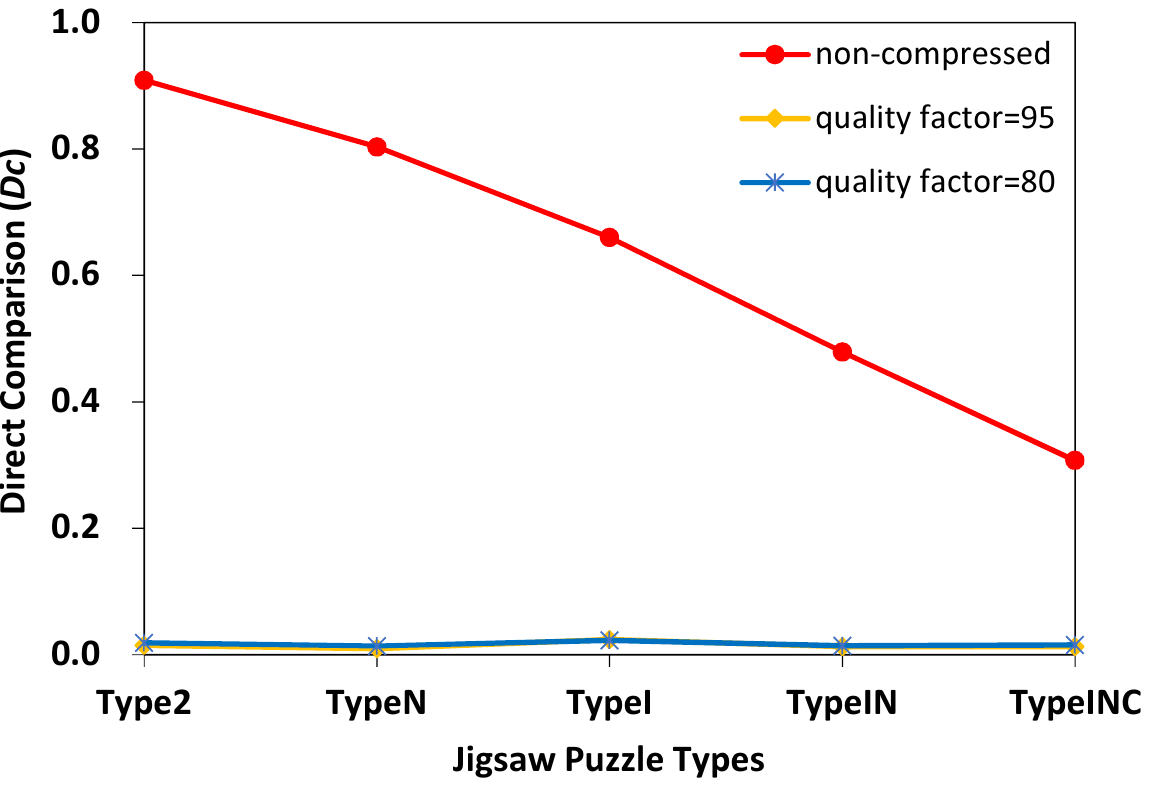}
          \hspace{3cm} {(a) Direct comparison ($Dc$) }
        \end{center}
      \end{minipage}
      \begin{minipage}{0.33\hsize}
        \begin{center}
          \includegraphics[width=5.82cm]{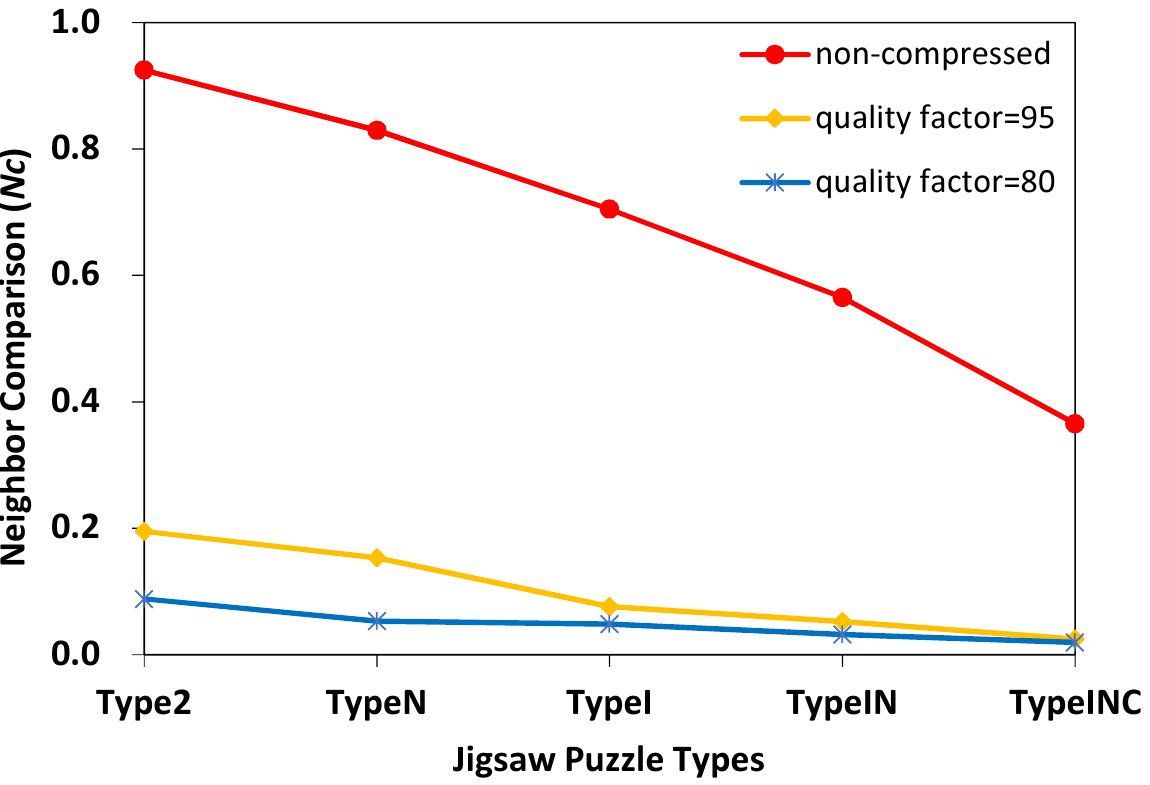}
          \hspace{3cm}  {(b) Neighbor comparison ($Nc$)}
        \end{center}
      \end{minipage}
            \begin{minipage}{0.33\hsize}
        \begin{center}
          \includegraphics[width=5.82cm]{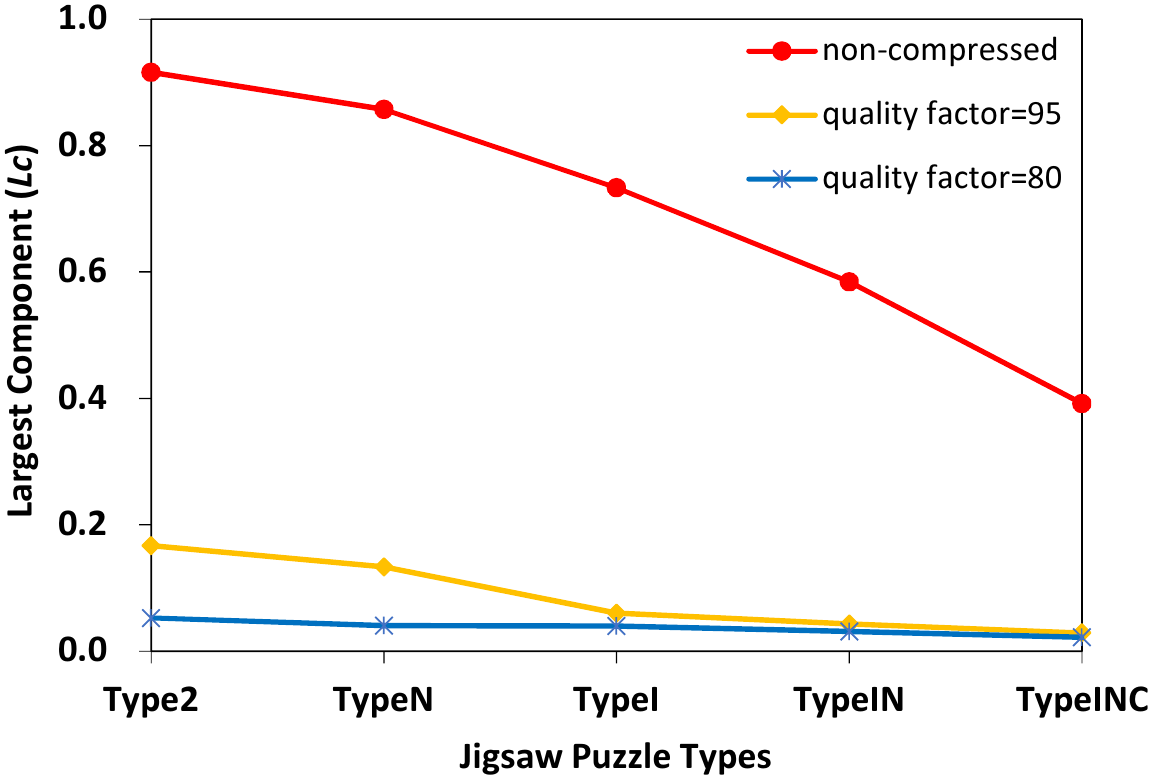}
          \hspace{3cm} {(c) Largest component ($Lc$)}
        \end{center}
      \end{minipage}
    \caption{Evaluation of the encrypted images including JPEG distortion
    against the jigsaw puzzle solver$\ (n=315$, $B_x \times B_y =$ $32 \times
    32$). The average of 20 images was evaluated.}
    \label{fig:dcnclc}
        \end{center}
\end{figure*}

 \begin{figure*}[t]
 \vspace{-0.5cm}
  \begin{center}
    \label{fig:puzzle}
      \begin{minipage}{0.24\hsize}
        \begin{center}
          \includegraphics[width=3.5cm]{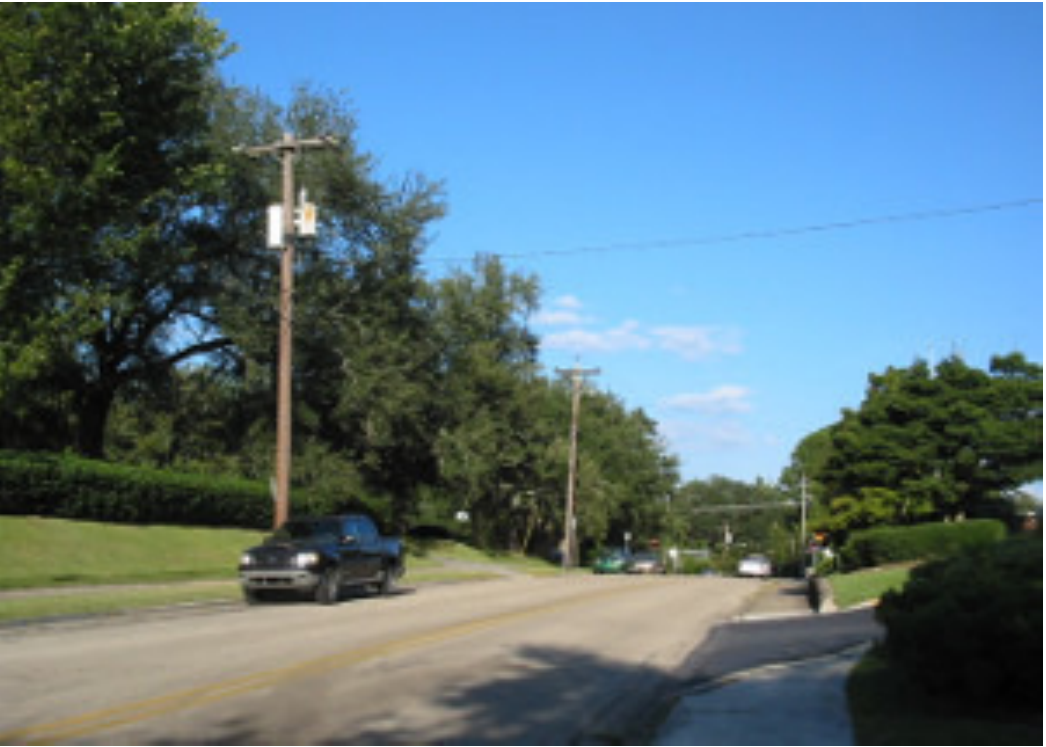}
          \hspace{3cm}{(a) Ordinary image\\}
          \vspace{0.7cm}
        \end{center}
      \end{minipage}
      \begin{minipage}{0.24\hsize}
        \begin{center}
          \includegraphics[clip, width=3.5cm]{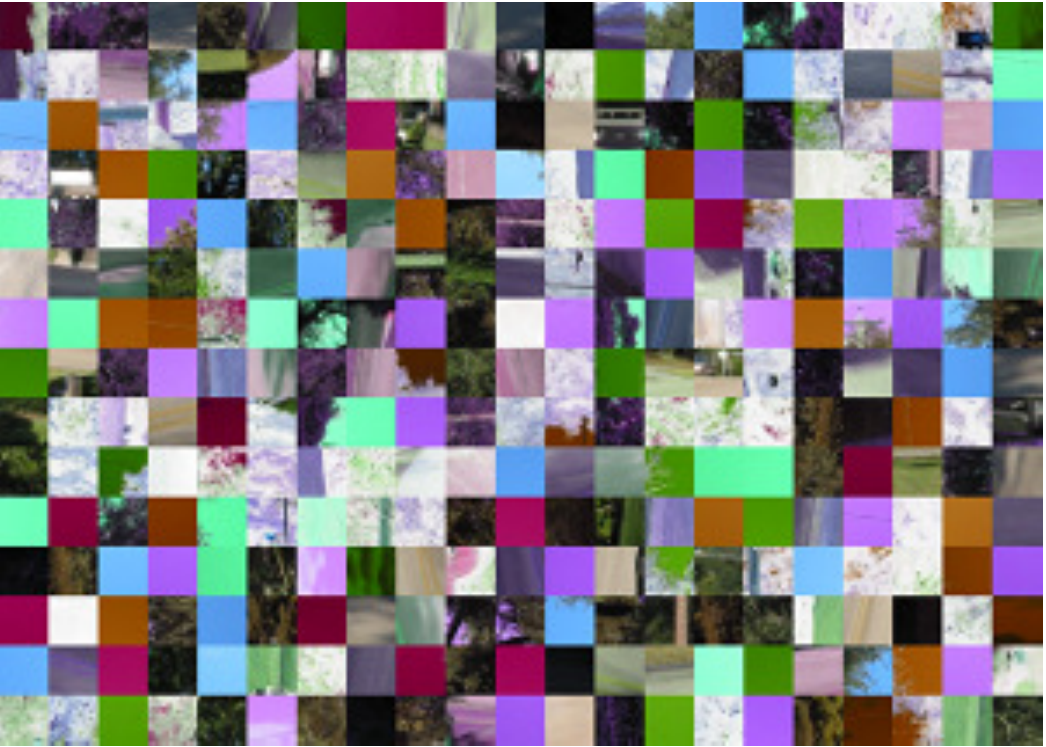}
          \hspace{3cm} {(b) Type INC puzzle\\}
          \vspace{0.7cm}
        \end{center}
      \end{minipage}
      \begin{minipage}{0.24\hsize}
        \begin{center}
          \includegraphics[clip, width=3.5cm]{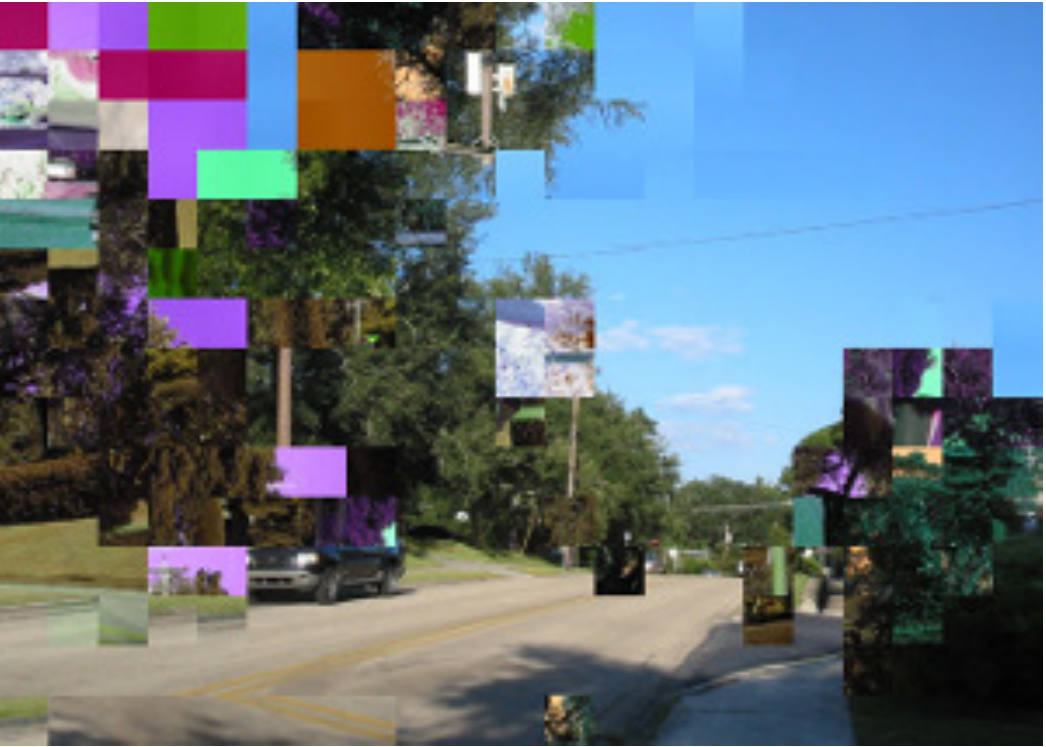}
          \hspace{3cm} {(c) Solved puzzle \\(TypeINC, non-compressed)\\
          \hspace{-4mm}$Dc=0.4,Nc=0.4,Lc=0.4$}
        \end{center}
      \end{minipage}
      \begin{minipage}{0.24\hsize}
        \begin{center}
          \includegraphics[clip, width=3.5cm]{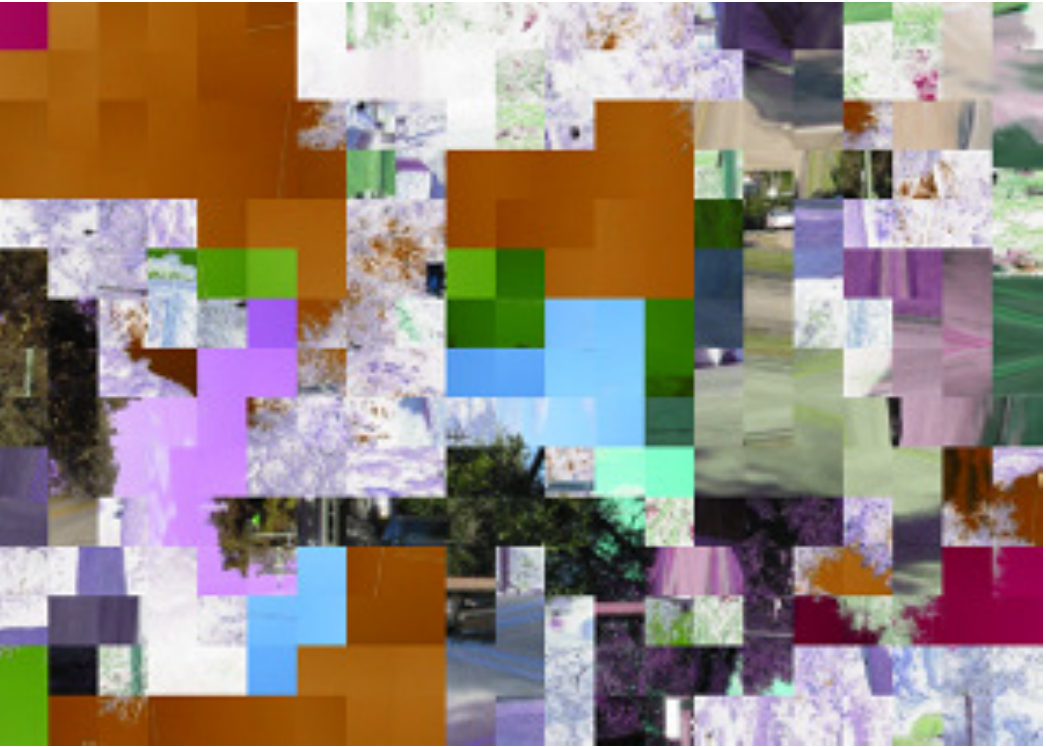}
          \hspace{3cm} {(d) Solved puzzle (TypeINC, JPEG compressed with Q=95)\\ \hspace{-4mm}$Dc=0,Nc=0,Lc=0$}
        \end{center}
      \end{minipage}
  \end{center}
    \caption{Examples of encrypted images and assembled images$ (n=315$, $B_x \times B_y =$ $32 \times 32$)}
    \label{fig:typeINC}
 \end{figure*}

\subsection{Experimental Result}
Figure\,\ref{fig:dcnclc} shows the scores of images assembled by the extended jigsaw puzzle solver\cite{CHUMAN2017ICASSP} discussed in Sec.\,\ref{sec:new}.
As shown in Fig.\,\ref{fig:dcnclc}, the scores of Type INC puzzles were slightly high as $L_{c}=0.392$ if they did not include any compression distortion.
On the other hand, considering the encrypted images including JPEG artifact, the scores become much lower.
For example, the scores of Type INC puzzles with JPEG artifact are very low as $L_{c}=0.029$, even when the high quality factor($Q=95$) in the JPEG compression was used.
Figures \,\ref{fig:typeINC}(c) and \ref{fig:typeINC}(d) show the example of assembled Type INC puzzles, where Fig.\,\ref{fig:typeINC}(d) was affected by JPEG compression and Fig.\,\ref{fig:typeINC}(b) was encrypted images generated from Fig.\,\ref{fig:typeINC}(a).
As well as Type INC puzzles with distortion, the scores of Type 2 puzzles which include only scrambled pieces and rotated ones, were very low as $L_{c}=0.053(Q=80)$.
It is confirmed that only few JPEG distortion makes puzzle solvers more difficult to assemble.

\section{Conclusion}
In this paper, the safety of the block-scrambling based image encryption schemes for EtC systems was discussed.
Also, we evaluated the performances of jigsaw puzzles with JPEG distortion.
We focused on jigsaw puzzle solvers as one of attack methods on the encryption, and regarded blocks of an encrypted image as pieces of a jigsaw puzzle, although the safety has been evaluated so far on the size of the key space, assuming the brute-force attacks.
In the simulations, it was shown that encrypted images including JPEG distortion are strong robustness against jigsaw puzzle solver attacks, in addition, the combination of each encryption step make assembling the images difficult.
{\tiny
\bibliographystyle{ieicetr}
}

\end{document}